# Modification to the Van der Waals Equation of State*

Jianxiang Tian[†]   Yuanxing Gui[#]

*Department of Physics, Dalian University of Technology*
*Dalian 116024, P.R.China*

**Abstract**

In this paper, we modify the VDW equation of state by adding a temperature factor to it. As a result, we give out a good phase diagram and the correlation of the reduced pressure and the reduced temperature when a balanced liquid-gas coexistence canonical argon-like system is considered.

*Supported by the National Natural Science Foundation of China under Grant No.10275008.
[†] Email address: lanmanhuayu@yahoo.com.cn
[#] Email address: guiyx@dlut.edu.cn

## 1   INTRODUCTION

Much attention has been paid in recent years to the hard core Yukawa (HCY) potential as a model for the pair interactions of fluids [1]. The liquid state theories such as the Mean Spherical Approximation (MSA) [2] and the Self Consistent Ornstein-Zernike Approximation (SCOZA) are proposed. Recent studies of the HCY fluid can be found in [2, 3] and references therein. Sutherland potential is a hard-core potential, too. Sutherland potential is written as

$$u = \begin{cases} \infty, & r < \sigma; \\ -\varepsilon\left(\dfrac{\sigma}{r}\right)^6, & r \geq \sigma. \end{cases} \quad (1)$$

Van der Waals mentioned out his famous equation of state in 1873 by adding an attraction term and a dispersion term to the ideal equation of state. The authors in Ref. (4) got the canonical partition function corresponded with the VDW equation of state. This canonical partition function is gotten by integrating under Sutherland potential. In PART 2 of this paper, we apply this canonical partition function to get the VDW equation of state, the chemical potential and their reduced forms. Then we study the case of the balanced liquid-gas coexistence argon-like canonical system. The theoretic data of the reduced density of the gas, the reduced density of the liquid, the reduced temperature and the reduced pressure under critical temperature are gotten and illustrated in figures. In PART 3, we modify the VDW equation of state by adding a temperature factor to it. As a result, the addition of this temperature factor does not affect the theoretic data of the reduced densities in the case we study. Worthwhile, it enhances the accuracy of the reduced temperature and the reduced pressure. Then a good diagram is given. So does the correlation between the reduced pressure and the reduced temperature.

## 2 PREPARATIVE WORK

### 2.1 The Canonical Partition Function

The authors in Ref. (4) offered the appropriate canonical ensemble partition function for their model [4]

$$Q = \frac{1}{N!\lambda^{3N}} \exp(-\beta N \phi/2)(V_f)^N$$

with $V_f = V - Nb$, $\lambda = \frac{h}{(2\pi m k_B T)^{1/2}}$. $k_B$ is the Boltzmann constant. The energy $\phi$ is given by [4]

$$\phi = -\int_\sigma^\infty \varepsilon \left(\frac{\sigma}{r}\right)^6 \frac{N}{V} 4\pi r^2 dr = -4\pi\varepsilon\sigma^3 N/V/3 \qquad (2)$$

Thus the canonical partition function is written as

$$Q = \frac{1}{N!\lambda^{3N}} \exp(\beta N 4\pi\sigma^3 N/V/6)(V_f)^N$$
$$= \frac{1}{N!\lambda^{3N}} \exp(\beta a N^2/V)(V - Nb)^N, \qquad (3)$$

with $a = 2\pi\varepsilon\sigma^3/3$.

## 2.2 The Equation of State

We get the equation of state as follows

$$P = k_B T \left(\frac{\partial \ln Q}{\partial V}\right)_{N,T} = \frac{Nk_B T}{V - Nb} - an^2. \qquad (4)$$

$n$ is the particle number density defined by $n = V/N$. Eq. (4) can also be expressed as

$$(P + an^2)(V - Nb) = Nk_B T \qquad (5)$$

$$\left(P + a\frac{1}{v^2}\right)(v - b) = k_B T \qquad (6)$$

This is just the VDW equation of state.

Here we sign the pressure of the gas $P_1$, the pressure of the liquid $P_2$, the chemical potential of the gas $\mu_1$, the chemical potential of the liquid $\mu_2$, the critical temperature $T_c$, the critical pressure $P_c$, the critical particle number density $n_c$, the particle number density of the gas $n_1$, the particle number density of the liquid $n_2$, the reduced temperature $T^*$, the reduced particle number density of the gas $n_1^*$, the reduced particle number density of the liquid $n_2^*$, the reduced pressure of the gas $P_1^*$, the reduced pressure of the liquid $P_2^*$, the particle number of the gas $N_1$, the particle number of the liquid $N_2$, the volume of the gas $V_1$, and the volume of the liquid $V_2$. The relations between them are

$$T^* = T/T_c, \qquad (7)$$
$$n_1 = N_1/V_1, \qquad (8)$$
$$n_2 = N_2/V_2, \qquad (9)$$
$$n_1^* = n_1/n_c, \qquad (10)$$
$$n_2^* = n_2/n_c, \qquad (11)$$
$$P_1^* = P_1/P_c, \qquad (12)$$
$$P_2^* = P_2/P_c. \qquad (13)$$

At critical point, the function $P = P(V)$ has such qualities as

$$\frac{\partial P}{\partial V}|_{T_c} = 0, \tag{14}$$

$$\frac{\partial^2 P}{\partial V^2}|_{T_c} = 0. \tag{15}$$

Thus we get the critical data by solving Eq. (14) and Eq. (15). They are

$$n_c = \frac{1}{3b}, \tag{16}$$

$$k_B T_c = \frac{8a}{27b}, \tag{17}$$

$$P_c = \frac{a}{27b^2}, \tag{18}$$

$$\frac{P_c}{n_c k_B T_c} = 3/8. \tag{19}$$

These are the critical data we are very familiar with.

Inputting Eq. (16-18) to Eq.(4), we get the reduced equation of state

$$P^* = \frac{8n^* T^*}{(3-n^*)} - 3n^{*2}. \tag{20}$$

## 2.3 The Chemical Potential

We can calculate the chemical potential as follows

$$\mu = -k_B T \left( \frac{\partial \ln Q}{\partial N} \right)_{V,T}$$

$$= -2an + k_B T \ln\left(\frac{n}{1-nb}\right) + \frac{nk_B Tb}{1-nb} - \frac{3k_B T}{2} \ln \frac{2\pi m k_B T}{h^2}. \tag{21}$$

Inputting Eq. (16-18) to Eq. (21), we can get the critical chemical potential. Then the reduced chemical potential is solved to be

$$\mu^* = \frac{A_1 + A_2}{A_3}, \tag{22}$$

where $A_1$, $A_2$ and $A_3$ are given by

$$A_1 = \left[ -n^* + \left( \ln \frac{n^*}{b(3-n^*)} \right) \frac{4T^*}{9} + \frac{4T^* n^*}{9(3-n^*)} \right], \tag{23}$$

$$A_2 = \left[ -\frac{2T^*}{3} \ln \frac{16\pi m a T^*}{27bh^2} \right], \tag{24}$$

$$A_3 = \left[ -1 + \left( \ln \frac{1}{2b} + \frac{1}{2} - \frac{3}{2} \ln \frac{16\pi m a}{27bh^2} \right) \frac{4}{9} \right]. \tag{25}$$

## 2.4 Balanced Liquid-gas Coexistence Canonical System

We consider the case of the balanced liquid-gas coexistence argon-like canonical system, which is a unit-two phases system. Here we do think the two phases are described by the same canonical partition function as Eq. (3). There are three balanced conditions, which must be satisfied when these two phases are balanced. They are thermal condition

$$T_1 = T_2, \tag{26}$$

dynamic condition

$$P_1 = P_2, \tag{27}$$

and phase condition

$$\mu_1 = \mu_2. \tag{28}$$

For Eq. (7-13), these three conditions can be expressed in the reduced unit as

$$T_1^* = T_2^*, \tag{29}$$

$$P_1^* = P_2^*, \tag{30}$$

$$\mu_1^* = \mu_2^*. \tag{31}$$

From Eq. (20), we have

$$P_1^* = \frac{8n_1^* T_1^*}{(3 - n_1^*)} - 3n_1^{*2}, \tag{32}$$

$$P_2^* = \frac{8n_2^* T_2^*}{(3 - n_2^*)} - 3n_2^{*2}. \tag{33}$$

From Eq. (22), we have

$$\mu_1^* = \frac{(A_1)_1 + (A_2)_1}{(A_3)_1}, \tag{34}$$

$$\mu_2^* = \frac{(A_1)_2 + (A_2)_2}{(A_3)_2}. \tag{35}$$

The solution of Eq. (29) and Eq. (30) is

$$T_1^* = T_2^* = T^* = \frac{9(n_1^* + n_2^*)(1 - n_c n_1^* b)(1 - n_c n_2^* b)}{8}. \tag{36}$$

The solution of Eq. (29) and Eq. (31) is

$$T_1^* = T_2^* = T^* = \frac{9(n_1^* - n_2^*)}{4\left[\ln\frac{n_1^*(1 - n_c n_2^* b)}{n_2^*(1 - n_c n_1^* b)} + \frac{n_c n_1^* b}{(1 - n_c n_1^* b)} - \frac{n_c n_2^* b}{(1 - n_c n_2^* b)}\right]}. \tag{37}$$

Thus we get a function $f(n_1^*, n_2^*) = 0$ easily from Eq. (36) and Eq. (37) with

$$f(n_1^*, n_2^*) = \frac{9(n_1^* + n_2^*)(1 - n_c n_1^* b)(1 - n_c n_2^* b)}{8}$$

$$- \frac{9(n_1^* - n_2^*)}{4\left[\ln\frac{n_1^*(1 - n_c n_2^* b)}{n_2^*(1 - n_c n_1^* b)} + \frac{n_c n_1^* b}{(1 - n_c n_1^* b)} - \frac{n_c n_2^* b}{(1 - n_c n_2^* b)}\right]} = 0 \tag{38}$$

We notice that $n_c b = \frac{1}{3}$, so $n_2^*$ can be gotten in the way of numerical computation by computer when an arbitrary $n_1^*$ is fixed. Substituting the data of $(n_1^*, n_2^*)$ to Eq. (36) or Eq. (37), we can get the data of $T^*$. Then substituting the data of $(n_1^*, n_2^*, T^*)$ to Eq. (20), we get the data of $P^*$. Thus we get the theoretic data of $(n_1^*, n_2^*, T^*, P^*)$. Some of them are illustrated in Table. (1). In Ref. (5), a method called polynomial approximation is introduced to get the relation of two variables. Here we adopt this method to get the relation of $(n_1^*, T^*)$ and the relation of $(n_2^*, T^*)$. The result is

$$n_1^* = \sum_{i=0}^{9} a_{1i} T^{*i}, \qquad (39)$$

$$n_2^* = \sum_{i=0}^{9} a_{2i} T^{*i}. \qquad (40)$$

Coefficients $\{a_{1i}\}$ and $\{a_{2i}\}$ are illustrated in Table. (2). And the relation of $\ln P^*$ versus $1/T^*$ is gotten by Matlab software to be

$$\ln P^* = 3.5204 - 3.5660/T^*. \qquad (41)$$

In 1945, E.A.Guggenheim collected the data of the balanced liquid-gas coexistence canonical system from experiments and gave out the correlation of $\rho_1^*$, $\rho_2^*$ and $T^*$ by the empirical equations [6] below

$$\rho_1^* = 1 + 0.75(1 - T^*) - 1.75(1 - T^*)^{1/3}, \qquad (42)$$

$$\rho_2^* = 1 + 0.75(1 - T^*) + 1.75(1 - T^*)^{1/3}. \qquad (43)$$

For

$$\rho_1^* = mn_1/mn_c = n_1/n_c = n_1^*, \qquad (44)$$

$$\rho_2^* = mn_2/mn_c = n_2/n_c = n_2^*, \qquad (45)$$

we have

$$n_1^* = 1 + 0.75(1 - T^*) - 1.75(1 - T^*)^{1/3}, \qquad (46)$$

$$n_2^* = 1 + 0.75(1 - T^*) + 1.75(1 - T^*)^{1/3}. \qquad (47)$$

In Eq. (42-45), $\rho_1^*$ is the reduced density of the gas and $\rho_2^*$ is the reduced density of the liquid. As far as an argon system is concerned, the inaccuracy of these equations is generally only one or two parts per thousand of $\rho_2^*$ when $T^* > 0.60 T_c$ [6]. So, it is acceptable to consider the data of $(T^*, n_1^*(T^*), n_2^*(T^*))$ from Eq. (46) and Eq. (47) as the experimental ones in this temperature region when the argon-like system is considered. In Eq. (44-45), $m$ is the mass of one particle. E.A.Guggenheim gave a numerical analytic result of the relation between the reduced temperature and the reduced pressure from experiments by equation [7]

$$P_e^* = \exp(5.29 - 5.31/T^*), \qquad (48)$$

which best fits the experimental data for argon except a tiny region near the critical point [7]. Thus it is acceptable to consider the data of $(P_e^*, T^*)$ from Eq. (48) as the experimental ones, too.

From Eq. (46-48), we get the experimental data of $(n_1^*, n_2^*, T^*, P_e^*)$. Some of them are illustrated in Table. (3). We compare these data with the theoretic ones we got above by illustrating them in figures. Fig. (1) is the curve of $n_1^*$ versus $n_2^*$. The real one is for the theoretic data from Eq. (38). The broken one is from Eq. (46-47). The figure tells us that the VDW equation of state offers good forecast to the correlation of $n_1^*$ and $n_2^*$ except the region near the critical point. Worthwhile the reduced temperature it offers is not so good. Fig. (2) is the phase diagram. The star curve comes from the theoretic data. The real curve is from Eq. (46-47). From Fig. (2) we see the relation between $n^*$ and $T^*$ deduced by the VDW equation of state is inaccurate. We also can find it from Table. (1) and Table. (3). Fig. (3) is the curve of $\ln P^*$ versus $1/T^*$. The real one comes from the data of Eq. (48). The star curve is from our theoretic data. We have noticed that the correlation between $\ln P^*$ and $1/T^*$ from Eq. (48) is expressed by a straight line and the one from the VDW equation of state is expressed by a straight line, too. It is wonderful. All the results above tell us that the VDW equation of state is good in a certain, but it is not right completely. So we try to modify it in PART 3 and go on studying the case of the balanced liquid-gas coexistence canonical system.

## 3 MODIFICATION

Now we modify the VDW equation of state by adding a temperature factor to it by the form of

$$P = \frac{Nk_B T}{V - Nb} - aT^S n^2. \tag{49}$$

$S$ is the power to be assured. From Eq. (3-4), we know this modification to the VDW equation of state is equal to such a modification to the canonical partition function as

$$Q = \frac{1}{N!\lambda^{3N}} \exp\left(\beta aT^S N^2 / V\right)(V - Nb)^N. \tag{50}$$

Then the chemical potential in this case can be gotten to be

$$\mu = -2aT^S n + k_B T \ln\left(\frac{n}{1-nb}\right) + \frac{nk_B Tb}{1-nb} - \frac{3k_B T}{2} \ln \frac{2\pi m k_B T}{h^2} \tag{51}$$

Solving Eq. (14-15), we get the critical data in this case. They are

$$n_c = \frac{1}{3b}, \tag{52}$$

$$k_B T_c^{1-S} = \frac{8a}{27b}, \tag{53}$$

$$P_c = T_c^S \frac{a}{27b^2}, \tag{54}$$

$$\frac{P_c}{n_c k_B T_c} = 3/8. \tag{55}$$

We are surprised to get these results. Then we get the reduced form of Eq. (49)

$$P^* = T^{*S}\left[\frac{8n^* T^{*(1-S)}}{(3-n^*)} - 3n^{*2}\right]. \tag{56}$$

And the reduced form of Eq. (51) can also be gotten to be

$$\mu^* = \frac{A_1^* + A_2^* + A_3^*}{A_4^*}, \tag{57}$$

with $A_1^*, A_2^*, A_3^*$ and $A_4^*$ are given by

$$A_1^* = T^{*S}\left[-n^* + \left(\ln \frac{n^*}{b(3-n^*)}\right)\frac{4T^{*1-S}}{9}\right], \tag{58}$$

$$A_2^* = T^{*S}\left[\frac{4T^{*1-S} n^*}{(3-n^*)9}\right], \tag{59}$$

$$A_3^* = T^{*S}\left[-\frac{2T^{*1-S}}{3} \ln \frac{16\pi m aT^* T_c^S}{27bh^2}\right], \tag{60}$$

$$A_4^* = \left[-1 + \left(\ln \frac{1}{2b} + \frac{1}{2} - \frac{3}{2} \ln \frac{16\pi m aT_c^S}{27bh^2}\right)\frac{4}{9}\right]. \tag{61}$$

Solving Eq. (29) and Eq. (30), we get the result below

$$T_1^{*(1-S)} = T_2^{*(1-S)} = T^{*(1-S)} = \frac{9(n_1^* + n_2^*)(1 - n_c n_1^* b)(1 - n_c n_2^* b)}{8}. \tag{62}$$

Solving Eq. (29) and Eq. (31), we get the results below

$$T_1^{*(1-S)} = T_2^{*(1-S)} = T^{*(1-S)}, \tag{63}$$

$$T^{*(1-S)} = \frac{9(n_1^* - n_2^*)}{4\left[\ln\frac{n_1^*(1 - n_c n_2^* b)}{n_2^*(1 - n_c n_1^* b)} + \frac{n_c n_1^* b}{(1 - n_c n_1^* b)} - \frac{n_c n_2^* b}{(1 - n_c n_2^* b)}\right]}. \tag{64}$$

Thus the correlation of $(n_1^*, n_2^*)$ is described by Eq. (38), too. At the base of the discussion in Part 2.4, we can get the theoretic data of $(n_1^*, n_2^*, T^*, P^*)$ when $S$ is fixed. Fig. (4) are the phase diagram with different values of $S$. Fig. (5) are the curve of $\ln P^*$ versus $1/T^*$ with the corresponded values of $S$. In order to get good theoretic data, we assure $S$ to be $-0.6393$ in the way of numerical computation by computer. Some of the theoretic data $(n_1^*, n_2^*, T^*, P^*)$ in this case are illustrated in Table. (4). Fig. (6) is the phase diagram when $S=-0.6393$. Fig. (7) is the curve of $\ln P^*$ versus $1/T^*$ when S=-0.6393. We can see the modification is effective in a certain. Then Eq. (49) reads

$$P = \frac{Nk_B T}{V - Nb} - aT^{-0.6393}n^2. \tag{65}$$

And its reduced form is

$$P^* = \frac{8n^* T^*}{(3 - n^*)} - 3T^{*-0.6393}n^{*2}. \tag{66}$$

# 4 DISCUSSION AND CONCLUSION

## 4.1 What We Have Done

In this paper, we modify the VDW equation of state by adding a temperature factor $T^S$ to it. As a result, a good phase diagram and the curve of $\ln P^*$ versus $1/T^*$ are given firstly when $S=-0.6393$. The phase diagram and the curve well approach the ones from experiments.

## 4.2 The Temperature Region

The temperature region we study is $1 > T^* > 0.6$ except a tiny region near the critical point, where Eq. (46-48) are in point.

## 4.3 About the VDW Equation of State

All the theoretic data are mainly illustrated in figures in this paper. From Fig.(1), we know that the VDW equation of state offers good theoretic forecast to the correlation between the reduced density of the liquid and the reduced density of the gas except the region near the critical point. At the same time the reduced temperature it offers is not very good, which can be seen from Fig. (2-3). From the data in Table. (1, 3), we can find this fact easily. For example, when $n_1^* = 0.0535$, $n_2^*$ from experiments is $2.3965$, $n_2^*$ from the VDW equation of state is

2.3313. The error is $|(2.3965\text{-}2.3313)/2.3965| = 0.0272$. But $T^*$ from experiments is 0.7000 and $T^*$ from the VDW equation of state is 0.5874. The error is $|(0.7000\text{-}0.5874)/0.7000|=0.1609$, which is larger than 0.0272.

## 4.4 About Eq.(49

Eq. (49) is a modification to the VDW equation of state. From Eq.(36-37,62-64) and Table.(1,4), we know Eq.(49) offers the same forecast to the correlation between the reduced density of the liquid and the reduced density of the gas as the one offered by the VDW equation of state. Fig. (4) indicates that the theoretic data of the reduced temperature delays on the value of $S$. Fig. (5) tells us that the reduced pressure delays on the value of $S$, too. But for that the correlation between the reduced density of the liquid and the reduced density of the gas it offers does not fit the experiments completely, it is impossible to find a proper value of $S$ to give the proper phase diagram and the proper correlation of $\ln P^*$ versus $1/T^*$ in this modification. What's more, the correlation between the reduced density of the liquid and the reduced density of the gas the VDW equation of state offers is not good in the region near the critical point. When $S=-0.6393$, Eq. (49) offers good forecast to the reduced temperature and the reduced pressure. But it is difficult to give a proper interpretation why such a temperature factor can enhance the accuracy of the theoretic data.

| $T^*$ | 0.4461 | 0.5259 | 0.5874 | 0.6434 | 0.6984 | 0.7549 | 0.8158 | 0.8861 |
|---|---|---|---|---|---|---|---|---|
| $n_1^*$ | 0.0106 | 0.0292 | 0.0535 | 0.0851 | 0.1266 | 0.1827 | 0.2627 | 0.3928 |
| $n_2^*$ | 2.5301 | 2.4223 | 2.3313 | 2.2408 | 2.1435 | 2.0322 | 1.8951 | 1.7023 |
| $P^*$ | 0.0123 | 0.0388 | 0.0767 | 0.1285 | 0.1981 | 0.2915 | 0.4193 | 0.6051 |

Table.1: Theoretic data from the VDW equation of State.

$a_{10}=-0.0054\times10^6$     $a_{11}=0.0743\times10^6$     $a_{12}=-0.4495\times10^6$     $a_{13}=1.5720\times10^6$

$a_{14}=-3.5022\times10^6$     $a_{15}=5.1549\times10^6$     $a_{16}=-5.0135\times10^6$     $a_{17}=3.1072\times10^6$

$a_{18}=-1.1138\times10^6$     $a_{19}=0.1760\times10^6$     $a_{20}=0.0050\times10^6$     $a_{21}=-0.0689\times10^6$

$a_{22}=0.4244\times10^6$  $a_{23}=-1.5121\times10^6$  $a_{24}=3.4349\times10^6$  $a_{25}=-5.1594\times10^6$

$a_{26}=5.1252\times10^6$  $a_{27}=-3.2474\times10^6$  $a_{28}=1.1911\times10^6$  $a_{29}=-0.1927\times10^6$

Table2: Coefficients for Eq. (39) and Eq. (40).

| $T^*$ | 0.6000 | 0.6500 | 0.7000 | 0.7500 | 0.8000 | 0.8500 | 0.9000 | 0.9500 |
|---|---|---|---|---|---|---|---|---|
| $n_1^*$ | 0.0106 | 0.0292 | 0.0535 | 0.0851 | 0.1266 | 0.1827 | 0.2627 | 0.3928 |
| $n_2^*$ | 2.5894 | 2.4958 | 2.3965 | 2.2899 | 2.1734 | 2.0423 | 1.8873 | 1.6822 |
| $P^*$ | 0.0284 | 0.0562 | 0.1007 | 0.1670 | 0.2599 | 0.3840 | 0.5434 | 0.7412 |

Table.3: Experimental data from Eq. (46-48).

| $T^*$ | 0.6112 | 0.6757 | 0.7228 | 0.7641 | 0.8033 | 0.8424 | 0.8832 | 0.9289 |
|---|---|---|---|---|---|---|---|---|
| $n_1^*$ | 0.0106 | 0.0292 | 0.0535 | 0.0851 | 0.1266 | 0.1827 | 0.2627 | 0.3928 |
| $n_2^*$ | 2.5301 | 2.4223 | 2.3313 | 2.2408 | 2.1435 | 2.0322 | 1.8951 | 1.7023 |
| $P^*$ | 0.0172 | 0.0518 | 0.0999 | 0.1643 | 0.2493 | 0.3610 | 0.5086 | 0.7085 |

Table.4: Theoretic data from the modification with S=-0.6393.

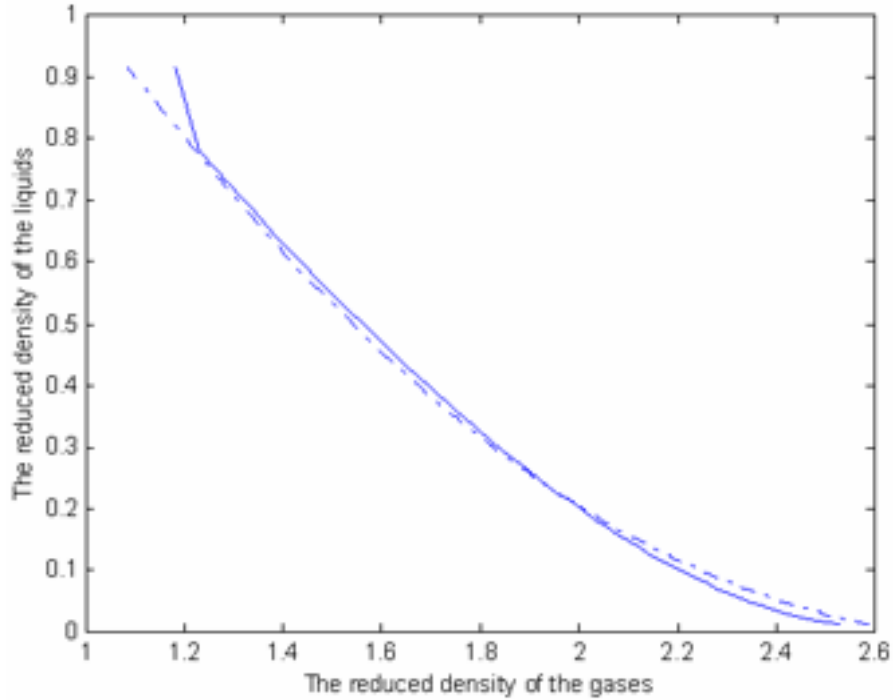

Figure 1: The curve of $n_1^*$ versus $n_2^*$. The broken line is from Eq. (46-47). The real one is from the VDW equation of State.

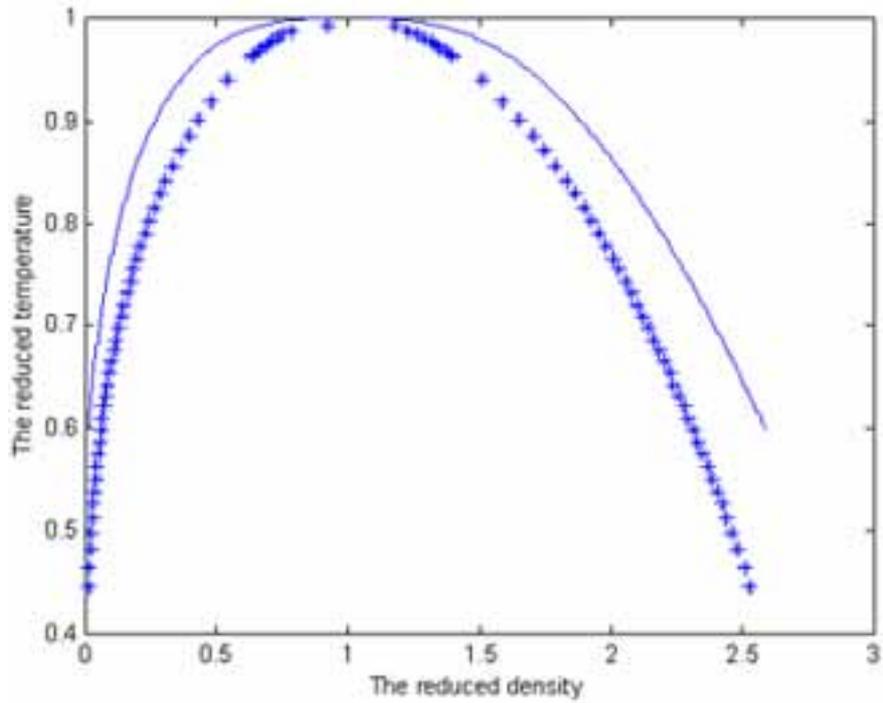

Figure 2: The phase diagram. The real line is from Eq. (46-47). The star one is from the VDW equation of state.

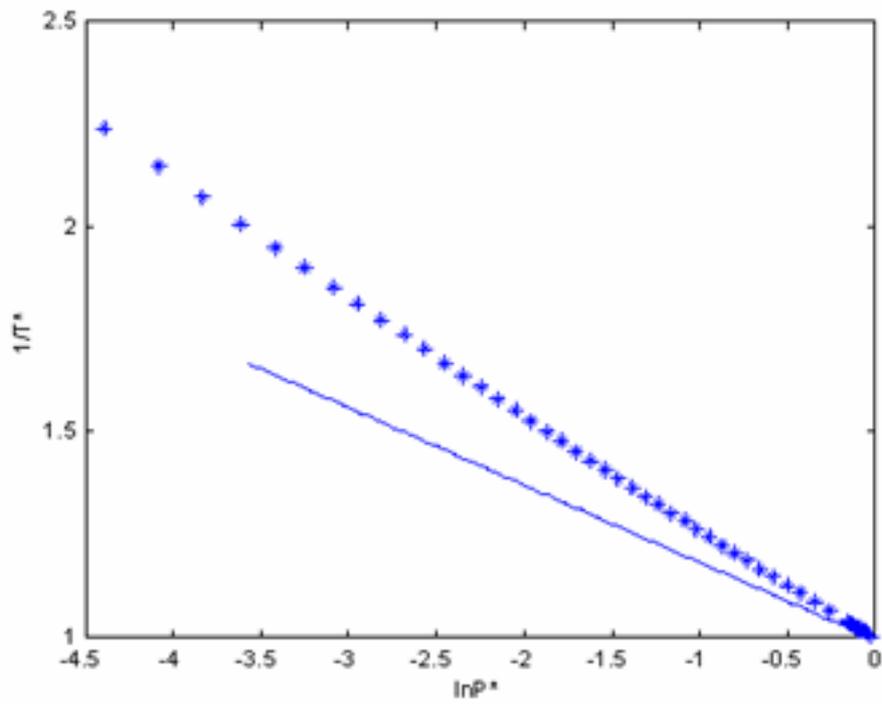

Figure 3: The curve of $\ln P^*$ versus $1/T^*$. The real one is from Eq. (48). The star one is from the VDW equation of state.

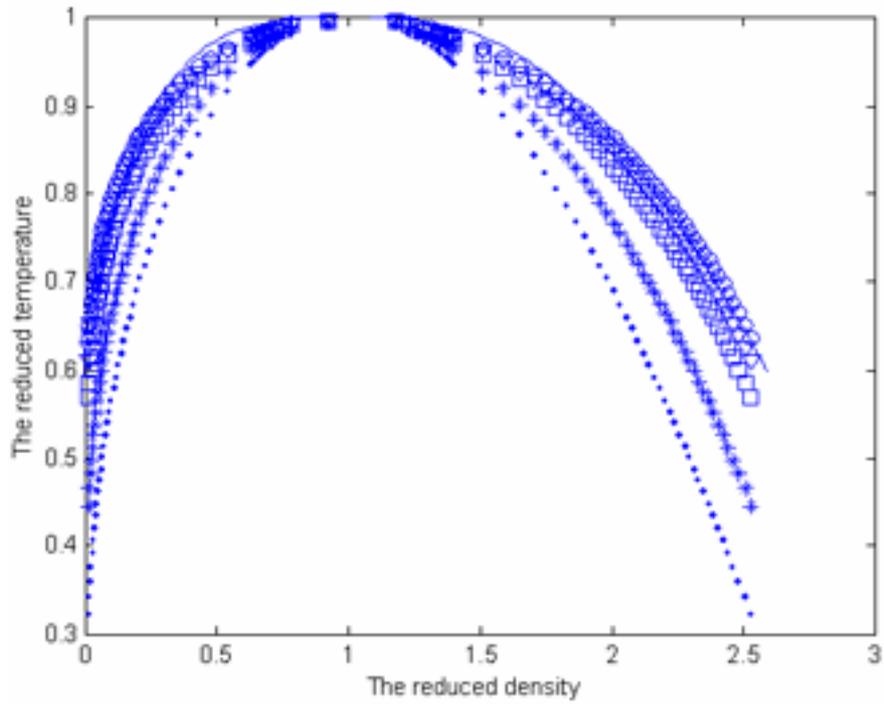

Figure 4: The phase diagram. The real line is from Eq. (46-47). The star one is from the VDW equation of state. The rest are all from the modification with different values of S: point—S=0.2857; square—S= -0.4286; triangle—S=-0.6393; circle—S=-0.7857.

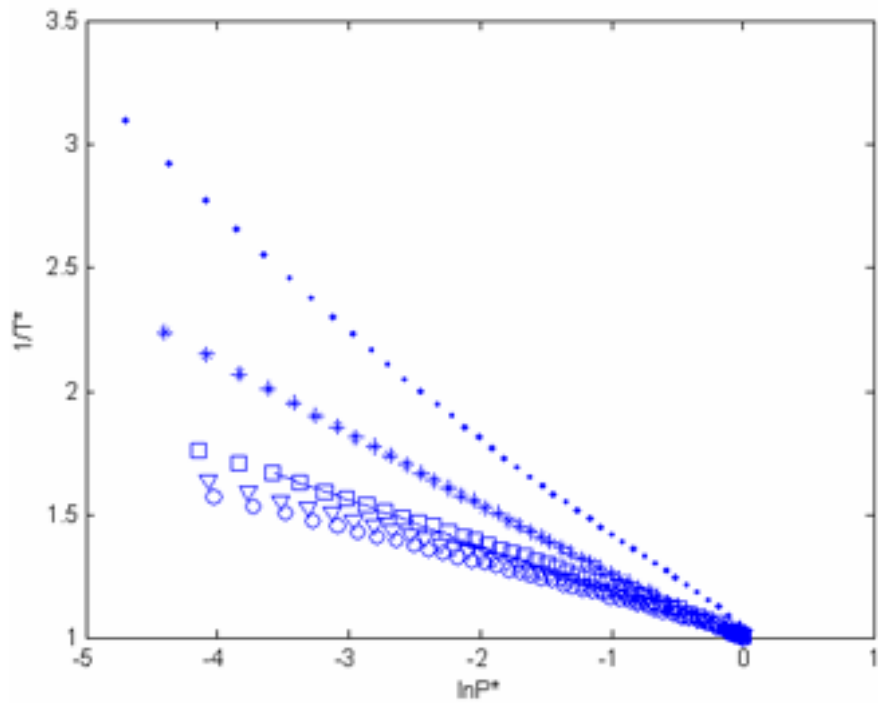

Figure 5: The curve of $\ln P^*$ versus $1/T^*$. The real one is from Eq. (48). The star one is from the VDW equation of state. The rest are all from the modification with different values of S. point—S=0.2857; square—S= -0.4286; triangle—S=-0.6393; circle—S=-0.7857.

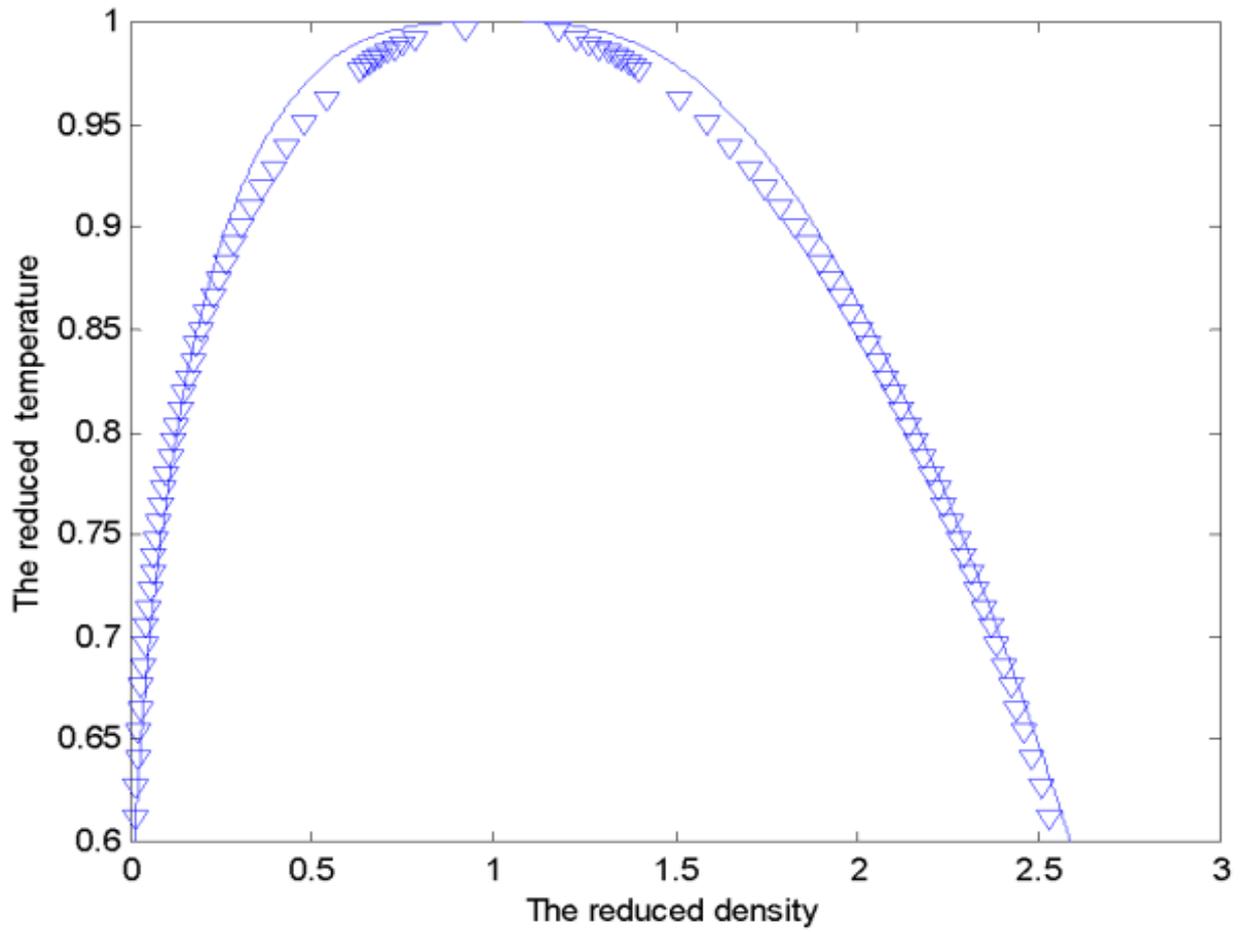

Figure 6: The phase diagram. The real line is from Eq. (46-47). The triangle one is from the modification with S=-0.6393.

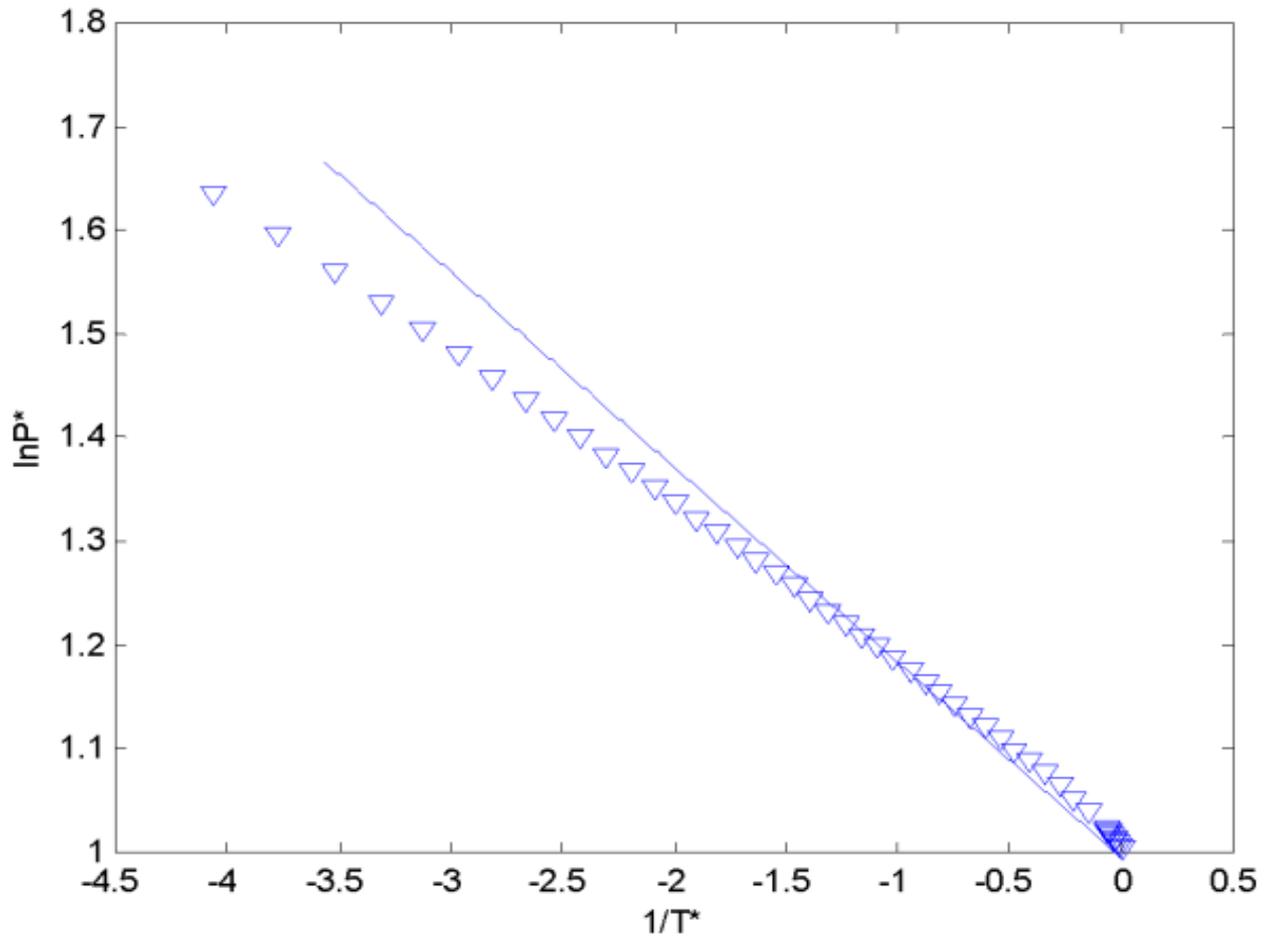

Figure 7: The curve of $\ln P^*$ versus $1/T^*$. The real one is from Eq. (48). The triangle one is from the modification with S=-0.6393 when the gas is considered.